\documentclass[conference]{IEEEtran}
\IEEEoverridecommandlockouts
\usepackage{cite}
\usepackage{amsmath,amssymb,amsfonts}
\usepackage{algorithmic}
\usepackage{graphicx}
\usepackage{textcomp}
\usepackage{xcolor}
\usepackage{url} 
\usepackage{hyperref}
\def\BibTeX{{\rm B\kern-.05em{\sc i\kern-.025em b}\kern-.08em
    T\kern-.1667em\lower.7ex\hbox{E}\kern-.125emX}}
\makeatletter
\newcommand{\linebreakand}{%
  \end{@IEEEauthorhalign}%
  \hfill\mbox{}\par%
  \mbox{}\hfill\begin{@IEEEauthorhalign}%
}
\makeatother
\begin{document}
\title{Analyzing RV32/RV64 Trade-offs for FreeRTOS Latency on 8-Stage RISC-V Soft Processors
}
\author{
\IEEEauthorblockN{
Hyunwoo Kang,
Geonwoo Yu,
Jongwon Kim,
Seungwoo You,
and Minchan Gil
}
\IEEEauthorblockA{
Department of System Semiconductor Engineering,
Sangmyung University\\
Cheonan, Republic of Korea\\
Email: \{hwctech1026, gu3205, kimjw033160,
youjm3535, ghilmc1019\}@gmail.com
}
}
\maketitle
\begin{abstract}
Although many commercial RISC-V platforms provide real-time operating system support, practical examples that explain how to enable a preemptive RTOS on a custom bare-metal RISC-V soft processor remain limited, leaving the interaction between processor microarchitecture, interrupt handling, and RTOS context switching difficult to understand from simple hardware implementation examples. This paper presents the design and evaluation of FreeRTOS on custom 8-stage RV32/RV64 RISC-V soft processors with machine-mode CSRs, a CLINT timer, trap and exception control logic, and the required context-switch path. Using this platform, we compare RV32 and RV64 under the same microarchitectural organization and firmware structure using two Rhealstone-derived latency microbenchmarks, task switching and task preemption, measured with the mcycle counter. RV64 requires 36.6\% and 17.7\% more cycles for task switching and preemption, respectively. Instruction-level analysis attributes this overhead to the doubled RV64 trap frame, wider pointer-based kernel data structures, and 64-bit scheduler priority handling. The RTL, firmware, and benchmark code are released as open source.
\end{abstract}


\section{Introduction}

RISC-V is an open instruction set architecture (ISA) that has seen rapid adoption in both industry and academia, making it an attractive target for custom processor research~\cite{mezger2022}.
FreeRTOS is a widely adopted open-source real-time operating system for microcontrollers~\cite{freertos_ug}, and its broad hardware support makes it a natural choice for validating custom processor designs.

While numerous studies have ported FreeRTOS to existing RISC-V platforms from a primarily software-oriented perspective, relatively few have examined the hardware--software integration challenges that appear when the target is a custom bare-metal RISC-V soft processor~\cite{mezger2022,neorv32,mattos2025,balas2021,jang2022,styger2019,batmaz2019,scheck2026}. 
Unlike vendor-supported or pre-validated platforms, where interrupt controllers, CSR behavior, and trap handling are already provided as part of the processor platform, a custom soft-core implementation requires these mechanisms to be verified together with the RTOS port. 
In particular, pipeline hazard behavior during interrupts, CLINT timer correctness, CSR read/write timing, and trap entry/exit sequencing directly affect the correctness of preemptive context switching.

Furthermore, prior RISC-V RTOS studies have primarily addressed RTOS porting, interrupt handling, context-switch optimization, or platform-level benchmarking~\cite{mezger2022,neorv32,mattos2025,balas2021,jang2022,styger2019,batmaz2019,scheck2026}. 
However, these studies do not explicitly isolate the effect of XLEN under the same microarchitectural organization and firmware structure. 
This distinction is relevant because the RISC-V ABI defines different data models for ILP32 and LP64, including 32-bit and 64-bit pointer widths, respectively~\cite{riscvpsabi}. 
In an RTOS implementation, this difference can influence pointer-based kernel data structures, while the architectural register width determines the amount of state saved and restored in trap and context-switch paths~\cite{freertos_riscv_port}. 
Although RV64 provides a larger address space and native 64-bit integer operations~\cite{riscvisa}, it may introduce additional RTOS latency through wider register save/restore operations and larger pointer-based data structures. 
Therefore, a controlled RV32/RV64 comparison provides useful guidance for selecting the appropriate XLEN in resource-constrained real-time embedded systems.

This paper makes three contributions. First, we describe the hardware extensions required to run FreeRTOS on a custom 8-stage pipelined RISC-V soft processor, including a finite state machine (FSM)-based trap controller, CLINT timer, and CSR infrastructure. Second, we port two Rhealstone-derived microbenchmarks (task switching and task preemption) to FreeRTOS and measure RTOS latency on both RV32 and RV64 configurations synthesized on the same FPGA platform. Third, we perform an instruction-level analysis of the compiled ELF binaries to identify the specific sources of the observed RV64 overhead. The complete RTL, FreeRTOS port, and benchmark source code are publicly
available in the RV-RTOS8 GitHub repository~\cite{rv-rtos8}.

\section{Processor Microarchitecture}
\subsection{8-Stage Pipeline Organization}

\begin{figure*}[t]
    \centering
    \includegraphics[width=\textwidth]{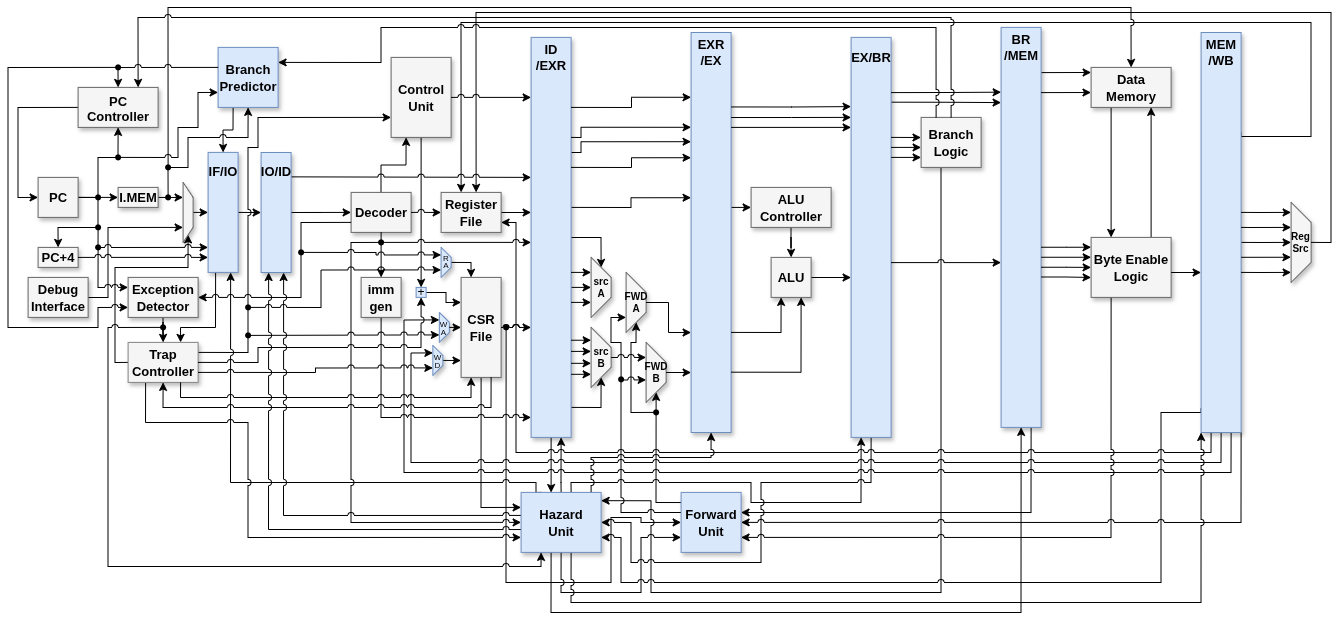}
    \caption{Block diagram of the 8-stage pipelined RISC-V processor core.}
    \label{fig:core}
\end{figure*}

The processor uses an 8-stage pipeline consisting of IF, IO, ID, EXR, EX, BR, MEM, and WB stages. The IO stage absorbs the one-cycle latency of synchronous BRAM instruction memory and provides an early branch-prediction point using a 2-bit saturating counter. The EXR stage separates forwarding and operand selection from ALU execution, reducing the EX-stage critical path. The BR stage verifies branch predictions and generates flush and PC-correction signals. The same pipeline organization is used for both RV32IM and RV64IM; the \texttt{XLEN} parameter controls datapath, register-file, and memory-interface widths. The 8-stage organization accommodates FPGA BRAM latency and critical-path constraints while providing a controlled platform for validating precise trap handling in a deep pipeline and comparing RV32/RV64 RTOS latency under an identical microarchitecture.
Fig.~\ref{fig:core} shows the overall processor structure.

\subsection{FSM-Based Trap Controller}
The deeper pipeline requires careful management of in-flight instructions during trap entry and return. We designed an FSM-based trap controller with 15 states that coordinates pipeline draining, CSR state preservation, and handler dispatch.

\subsubsection{Trap Detection and Pipeline Draining}
The \texttt{ExceptionDetector} module monitors multiple pipeline stages simultaneously and assigns priority to concurrent trap sources: MEM-stage exceptions (misaligned load/store) take highest priority, followed by EX2, EX, EXR, and ID-stage synchronous exceptions (\texttt{ECALL}, \texttt{EBREAK}, \texttt{MRET}, misaligned branch target, \texttt{FENCE.I}), with asynchronous timer interrupts at the lowest priority. A timer interrupt is qualified only when \texttt{mstatus.MIE}, \texttt{mie.MTIE}, and the CLINT \texttt{timer\_interrupt} signal are all asserted, and no synchronous exception is currently being handled.

When the Exception Detector asserts a trap, the FSM leaves \texttt{idle} and asserts \texttt{trap\_done\,=\,0}, which stalls the IF stage to halt new instruction fetch. For synchronous traps (\texttt{ECALL}, timer interrupt), the FSM enters a three-cycle pipeline draining sequence (\textsc{mem\_standby} $\to$ \textsc{wb\_standby} $\to$ \textsc{rtre\_standby}) to ensure all preceding instructions complete write-back before modifying CSR state. Load/store misalignment exceptions trigger dedicated flush signals to immediately discard the faulting instruction's effects.

\subsubsection{CSR State Preservation}
After the pipeline is drained, the FSM writes the faulting or interrupted PC to \texttt{mepc}. For timer interrupts, the mepc source is selected via a priority chain that searches from the deepest pipeline stage (WB) backward to IF, selecting the first non-zero PC to identify the most recently committed instruction point. The FSM then advances to \textsc{write\_mepc} $\to$ \textsc{write\_mcause}, where the appropriate cause code is recorded.

During the \textsc{write\_mepc} cycle, the CSR file performs the \texttt{mstatus} MIE/MPIE transition mandated by the RISC-V Privileged Specification~\cite{riscv_priv}: \texttt{MPIE\,$\leftarrow$\,MIE} and \texttt{MIE\,$\leftarrow$\,0}. This is implemented as a hardware-automatic operation triggered by the \texttt{trapped\_latch} signal, ensuring atomicity without software intervention and preventing nested interrupts during trap entry.

\subsubsection{Handler Dispatch and Return}
In the \textsc{write\_mcause} state, the FSM reads \texttt{mtvec} to obtain the trap handler base address. In the subsequent \textsc{read\_mtvec} $\to$ \textsc{goto\_mtvec} states, this address is injected as the next PC, while \texttt{pth\_done\_flush} simultaneously flushes the IF, IO, ID, and EXR stages. The \texttt{trap\_done} signal is reasserted, releasing the pipeline stall and allowing the processor to begin fetching from the trap handler.

On \texttt{MRET}, the FSM reads \texttt{mepc} and traverses five states (\textsc{read\_mepc} through \textsc{goto\_mret}) to propagate the restored PC through the 8-stage pipeline and flush stale instructions. In the final \textsc{goto\_mret} state, the \texttt{mret\_executed} signal triggers the CSR file to restore interrupts: \texttt{MIE\,$\leftarrow$\,MPIE}, \texttt{MPIE\,$\leftarrow$\,1}, as specified in the RISC-V Privileged Specification~\cite{riscv_priv}.

\subsection{CLINT Timer and CSR Infrastructure}
The Core Local Interruptor (CLINT) is a timer peripheral originally specified by SiFive~\cite{sifive_clint} and widely adopted in RISC-V SoC designs, including the FreeRTOS RISC-V port. Our CLINT module implements the \texttt{mtime} and \texttt{mtimecmp} registers as 64-bit MMIO-mapped registers at the standard addresses (\texttt{0x0200\_0000} and \texttt{0x0200\_0008}). A clock divider converts the 100\,MHz system clock into a 1\,kHz tick that increments \texttt{mtime} every 100{,}000 cycles. The 1\,kHz tick frequency matches the FreeRTOS \texttt{configTICK\_RATE\_HZ\,=\,1000} setting, providing 1\,ms scheduling granularity as recommended by the FreeRTOS documentation~\cite{freertos_ug}. The timer interrupt output is asserted when $\texttt{mtime} \geq \texttt{mtimecmp}$.

The CSR file implements the machine-mode registers required by the RISC-V Privileged Specification~\cite{riscv_priv}: \texttt{mstatus} (with hardware-driven MIE/MPIE transitions as described above), \texttt{mie} (read/write; bit~7 controls MTIE), \texttt{mip} (read-only; bit~7 reflects the CLINT \texttt{timer\_interrupt} signal), \texttt{mepc}, \texttt{mcause}, \texttt{mtvec}, \texttt{mscratch}, \texttt{misa}, \texttt{mvendorid}, \texttt{marchid}, and \texttt{mhartid}. The \texttt{mcycle} and \texttt{minstret} performance counters are also implemented, with \texttt{mcycle} incrementing every clock cycle and \texttt{minstret} incrementing on each non-NOP instruction retirement, enabling cycle-accurate benchmarking.

\section{FreeRTOS Porting and Benchmark Integration}

\subsection{FreeRTOS Port Layer}
FreeRTOS was ported to the custom processor using the official RISC-V FreeRTOS port as a starting point. The port layer consists of the trap handler (\texttt{portASM.S}), the configuration header (\texttt{FreeRTOSConfig.h}), a custom linker script, and startup code (\texttt{crt0.S}). The trap handler follows the standard FreeRTOS RISC-V convention: on entry, the full register context (all 31 general-purpose registers, \texttt{xCriticalNesting}, and \texttt{mstatus}) is saved to the current task's stack, and the stack pointer is stored into the TCB. After calling \texttt{xTaskIncrementTick} and conditionally \texttt{vTaskSwitchContext}, the scheduler selects the next task, and the corresponding context is restored from the new task's stack.

A key difference between the RV32 and RV64 ports lies in the trap frame size. The RV32 trap handler allocates 124\,bytes of stack space (31 four-byte words) and uses \texttt{sw}/\texttt{lw} instructions for context save/restore. The RV64 handler allocates 248\,bytes (31 eight-byte doublewords) and uses \texttt{sd}/\texttt{ld} instructions. While the instruction count is identical (31 stores on entry, 31 loads on exit), each RV64 memory access transfers twice the data, increasing memory bandwidth consumption per context switch.

\subsection{Rhealstone Benchmark Porting}
Rhealstone is an RTOS performance benchmark metric originally proposed to quantify
multiple real-time operating-system activities, including task switching, preemption,
interrupt latency, semaphore shuffling, deadlock breaking, and intertask communication
performance~\cite{rhealstone}.
Among the publicly available implementations examined in this work, the RTEMS
implementation provided the most directly reusable source-level realization of these
tests~\cite{rtems_rhealstone}.
Using this implementation as a reference, we ported the task-switching and
task-preemption tests to the FreeRTOS API and used them as Rhealstone-derived latency
microbenchmarks.
Both tests share a common parameter $N{=}1{,}000$ benchmark rounds and use the
\texttt{mcycle} CSR for cycle-accurate timing.
Loop overhead measured separately without actual task switching or preemption is
subtracted in each test to isolate the kernel cost.

\textit{Task switching} measures the average context-switch latency between two
equal-priority tasks that repeatedly yield to each other.
The RTEMS yield primitive was replaced with FreeRTOS \texttt{taskYIELD()}, which
triggers an \texttt{ECALL}-based synchronous context switch.
Each of the two tasks invokes \texttt{taskYIELD()} $N$ times, producing
$2N{=}2{,}000$ context-switch events; the total elapsed cycles are divided by $2N$.

\textit{Task preemption} measures the latency of a full preemption round trip:
the low-priority task resumes a suspended high-priority task via
\texttt{vTaskResume()}, is preempted, and regains control after the high-priority
task calls \texttt{vTaskSuspend()} on itself.
The RTEMS suspend/resume primitives were replaced with the corresponding FreeRTOS
calls.
The total elapsed cycles over $N$ such round trips are divided by $N$.

\section{Experimental Results and Analysis}

\subsection{Experimental Setup}

Both the RV32IM and RV64IM configurations were synthesized on a Digilent Nexys Video board with an AMD Xilinx Artix-7 XC7A200T-1SBG484C FPGA, operating at 100\,MHz. Synthesis used Vivado 2025.2 with \texttt{Performance\_Explore} and \texttt{Performance\_ExplorePostRoutePhysOpt} strategies. 
The experiments used FreeRTOS V11.1.0+, compiled with \texttt{riscv64-unknown-elf-gcc} 15.2.0 and \texttt{-Os}, using the \texttt{ilp32} ABI for RV32IM and \texttt{lp64} for RV64IM.
The SoC integrates PS/2 keyboard input, HDMI text-mode display, and UART serial output for benchmark result reporting. 


\subsection{Resource Utilization}

To isolate the hardware cost of the RTOS-capable processor subsystem, resource utilization was measured using a minimal FreeRTOS SoC configuration. 
This configuration includes the processor core, CLINT timer, and MMIO interface, with the CLINT and MMIO implementation kept identical between RV32IM and RV64IM. It excludes instruction/data memories, UART TX, PS/2 keyboard logic, HDMI display logic, VRAM, font ROM, and video peripherals. Table~\ref{tab:resource} shows the post-implementation resource utilization and estimated on-chip power.

This resource increase is expected because the RV64 configuration widens the datapath and architectural state while preserving the same 8-stage pipeline organization. The result highlights the area--latency trade-off of selecting RV64 for RTOS-capable embedded soft processors.

\begin{table}[t]
\caption{Resource Utilization of Minimal FreeRTOS SoC}
\label{tab:resource}
\centering
\begin{tabular}{lccc}
\hline
\textbf{Configuration} & \textbf{LUTs} & \textbf{FFs} & \textbf{Estimated Power} \\
 & & & \textbf{(W)} \\
\hline
RV32IM Minimal SoC & 3{,}264 & 2{,}480 & 0.448 \\
RV64IM Minimal SoC & 8{,}349 & 6{,}937 & 0.469 \\
\hline
\end{tabular}
\end{table}

\subsection{Benchmark Results}

Table~\ref{tab:rhealstone} summarizes the Rhealstone results. In task switching, RV32 recorded 320 cycles/iteration (3.20\,$\mu$s), while RV64 recorded 437 cycles/iteration (4.37\,$\mu$s), a 36.6\% increase. In task preemption, RV32 measured 1{,}074 cycles/iteration (10.74\,$\mu$s) versus 1{,}264 for RV64 (12.64\,$\mu$s), a 17.7\% increase.

Task preemption exhibits a smaller relative overhead despite higher absolute cycle count. This is because preemption involves additional scheduler operations (priority comparison, task state transitions via \texttt{vTaskResume}/\texttt{vTaskSuspend}) whose instruction counts are comparable between RV32 and RV64, diluting the relative contribution of context-switch overhead.

\begin{table}[t]
\caption{Rhealstone Benchmark Results (FreeRTOS, 100\,MHz)}
\label{tab:rhealstone}
\centering
\begin{tabular}{lccc}
\hline
\textbf{Sub-benchmark} & \textbf{RV32} & \textbf{RV64} & \textbf{RV64} \\
 & \textbf{(cycles/iter)} & \textbf{(cycles/iter)} & \textbf{Overhead} \\
\hline
Task Switch  & 320   & 437   & +36.6\% \\
Task Preempt & 1{,}074 & 1{,}264 & +17.7\% \\
\hline
\end{tabular}
\end{table}

\subsection{Instruction-Level Analysis}

To identify the root causes of the RV64 overhead, we disassembled the compiled ELF binaries and performed a detailed comparative analysis of the context-switch critical path. Table~\ref{tab:instlevel} summarizes the key differences.

\begin{table}[t]
\caption{Instruction-Level Comparison of Context-Switch Path}
\label{tab:instlevel}
\centering
\begin{tabular}{lcc}
\hline
\textbf{Component} & \textbf{RV32} & \textbf{RV64} \\
\hline
Trap frame size                        & 124\,B (31$\times$4) & 248\,B (31$\times$8) \\
Save/restore instructions              & 62 (31+31) & 62 (31+31) \\
Bytes per save+restore                 & 248\,B & 496\,B \\
\texttt{vTaskSwitchContext} instrs     & 34 & 37 \\
Priority CLZ routine                   & \texttt{\_\_clzsi2} & \texttt{\_\_clzdi2} \\
\texttt{List\_t} size (per level)      & 20\,B & 40\,B \\
\texttt{pxReadyTasksLists} total       & 100\,B & 200\,B \\
Timer interrupt path instrs            & 25 & 18 \\
\hline
\end{tabular}
\end{table}

\subsubsection{Trap Frame Overhead}
The dominant source of the RV64 cycle increase is the doubled trap frame. Both configurations save and restore 31 general-purpose registers, \texttt{xCriticalNesting}, and \texttt{mstatus} (62 memory operations total). RV32 uses \texttt{sw}/\texttt{lw} (4\,bytes each; 248\,bytes per full save+restore pair), while RV64 uses \texttt{sd}/\texttt{ld} (8\,bytes each; 496\,bytes). On the target FPGA's BRAM-based memory subsystem without data cache, the doubled data volume translates directly to additional memory stall cycles. Since each task switch involves two full context switches (saving the old task and restoring the new one), the trap frame overhead is the single largest contributor to the measured cycle difference.

\subsubsection{Scheduler Priority Handling}
In \texttt{vTaskSwitchContext}, the scheduler resolves the highest ready priority via a count-leading-zeros (CLZ) operation on \texttt{uxTopReadyPriority}. RV32 calls \texttt{\_\_clzsi2} (32-bit, 76\,bytes), while RV64 calls \texttt{\_\_clzdi2} (64-bit, 64\,bytes). The RV64 path contains three additional instructions: \texttt{slli}/\texttt{srli} by 32 to zero-extend the 32-bit bitmask to 64 bits before the CLZ call, and \texttt{addiw} to correct the result for 32-bit semantics (37 vs.\ 34 total instructions).

\subsubsection{Kernel Data Structure Widening}
FreeRTOS's internal data structures (TCBs, list nodes, scheduler state variables) use pointer-width fields that scale with XLEN. \texttt{List\_t} grows from 20\,B to 40\,B, and \texttt{pxReadyTasksLists} (5 priority levels) from 100\,B to 200\,B. Scalar variables (\texttt{xTickCount}, \texttt{pxCurrentTCB}, \texttt{uxTopReadyPriority}) widen from 4 to 8\,bytes. During the \texttt{vTaskSwitchContext} list traversal, these wider \texttt{ld}/\texttt{sd} accesses accumulate additional memory cycles.

\subsubsection{Timer Interrupt Path (RV64 Advantage)}
The timer interrupt handling path is \textit{shorter} in RV64 (18 instructions) than in RV32 (25 instructions). RV32 must perform 64-bit \texttt{mtime}/\texttt{mtimecmp} arithmetic using register pairs with explicit carry propagation (\texttt{sltu} + \texttt{add}), and must write \texttt{mtimecmp} via a three-step atomic sequence (write \texttt{0xFFFFFFFF} to low word, write high word, then write low word) to prevent spurious timer interrupts from intermediate values~\cite{riscv_priv}. RV64 performs the same operation with a single \texttt{ld}, \texttt{add}, and \texttt{sd}. This advantage partially offsets RV64 overhead in the preemption benchmark, where the timer interrupt path is exercised on every preemptive context switch, and explains why the preemption overhead (17.7\%) is smaller than the task-switch overhead (36.6\%), where the \texttt{ECALL} path bypasses timer handling entirely.

\section{Conclusion}

This paper presented a FreeRTOS-capable 8-stage RV32/RV64 RISC-V soft processor with CLINT-based timer interrupts, machine-mode CSRs, and an FSM-based trap controller for pipeline-aware trap entry and return. 

The results show that RV64 requires 36.6\% more cycles for task switching and 17.7\% more cycles for task preemption than RV32. Instruction-level analysis indicates that the main sources of this overhead are the doubled RV64 trap frame, wider pointer-based kernel data structures, and additional scheduler instructions. Although RV64 reduces the timer-update path through native 64-bit arithmetic, this benefit only partially offsets the increased context-switch cost.

These results suggest that RV32 remains preferable for latency-sensitive embedded RTOS workloads when a 64-bit address space or native 64-bit arithmetic is not required. 
Future work will extend the evaluation to the remaining Rhealstone sub-benchmarks and investigate how a data-cache subsystem affects the RV32/RV64 latency trade-off.

\vspace{12pt}

\end{document}